\documentclass{aastex}
\usepackage{emulateapj5}
\usepackage{epsf}
\usepackage{natbib}

\input{psfig}
\newenvironment{onecolumnfigure}{
\def\@captype{figure}
\noindent\begin{minipage}{0.9999\linewidth}\begin{center}}
{\end{center}\end{minipage}\smallskip}

\font \bolditalics = cmmib10
\def\bx#1{\leavevmode\thinspace\hbox{\vrule\vtop{\vbox{\hrule\kern1pt
        \hbox{\vphantom{\tt/}\thinspace{\bf#1}\thinspace}}
      \kern1pt\hrule}\vrule}\thinspace}
\def \vc #1{{\textfont1=\bolditalics \hbox{$\bf#1$}}}

\def\thetag{{\vc \theta}}

\def\gammag{{\vc \gamma}}

\begin{document}

\def\ve{{\vec e}} \def\vl{{\vec l}} \def\vx{{\vec x}} \def\vk{{\vec
k}} \def\vgam{{\vec{\gamma}}} \def\mD{{\cal{D}}} \def\d{{\rm d}}
\def\i{{\rm i}} \def\veps{{\vec \epsilon}} \def\eps{\epsilon}
\def\gam{\gamma}

\shortauthors{L. Van Waerbeke et
al.} \shorttitle{Cluster Magnification}
\title{Magnification as a Probe of Dark Matter Halos at high redshift}
\author{L. Van Waerbeke$^{1}$, H. Hildebrandt$^{2}$, J. Ford$^{1}$, M. Milkeraitis$^{1}$}
\newcommand{\etal}{{\it et al. }} \newcommand{\beq}{\begin{equation}}
\newcommand{\eeq}{\end{equation}}

\begin{abstract}
We propose a new approach for measuring the mass profile and shape
of groups and clusters of galaxies, which uses lensing magnification
of distant background galaxies. The main advantage of lensing
magnification is that, unlike lensing shear, it relies on accurate
photometric redshifts only and not galaxy shapes, thus enabling the
study of the dark matter distribution with unresolved source
galaxies. We present a feasibility study, using a real population of
$z\ge 2.5$ Lyman Break Galaxies as source galaxies, and where,
similar to galaxy-galaxy lensing, foreground lenses are stacked in
order to increase the signal-to-noise. We find that there is an
interesting new observational window for gravitational lensing as a
probe of dark matter halos at high redshift, which does not require
measurement of galaxy shapes.
\end{abstract}

\bigskip

\section{Introduction}

Dark matter halos can be used as a probe of both cosmological
parameters and structure formation. Their statistical distribution
as a function of mass and redshift is a strong probe of dark energy
\citep{allen2004}, and a comparison of the halo mass versus galaxy
distribution within individual halos contains important clues about
the role of dark matter in the baryonic mass buildup
\citep{hoekstra2005}. Any study using dark matter halos,
statistically or individually, requires an estimate of the mass
profile or the halo mass. Unfortunately, there are still practical
difficulties with these measurements, particularly for high redshift
halos and certain mass range. High redshift clusters, for instance,
are dynamically young objects, making velocity dispersion and X-ray
temperature based mass estimates questionable. There is therefore a
vivid interest in finding a reliable mass calibration tool: one that
is unbiased and, as much as possible, independent of the dynamical
state of the halo. In fact, the mass calibration issue is central to
high redshift cluster searches, such as the SpARCS/SWIRE survey
\citep{muzzin2009} and Sunyaev-Zel'dovich surveys.

The most prevalent technique for estimating halo masses uses
gravitational lensing, and more specifically the tangential shear
measured from the shapes of background galaxies. The variation of
the strength of the shear from the halo center is an estimate of the
lens mass profile. Although this technique is formally unbiased and
provides a mass estimate independent of the halo dynamical state and
shape, it suffers from practical limitations. Accurate galaxy shape
measurement requires good pixel resolution and a small Point Spread
Function (PSF), which, for high redshift sources, is generally not
the case. Current ground-based surveys show that it is unlikely one
can reliably measure the shape of galaxies at redshifts higher than
$z\simeq 1.5$. This upper bound automatically sets a maximum
redshift

\vbox{ \vspace{0.5cm}\footnotesize \noindent $^1$~University of
British Columbia,
Vancouver, B.C. V6T 2C2, Canada\\
$^2$~Leiden Observatory, Leiden University, Niels Bohrweg 2, 2333CA
Leiden, The Netherlands \\}

\noindent for the lenses one can probe: roughly half of the source
redshift. One concludes that it is unlikely shear observations can
probe the dark matter distribution at redshifts higher than $z\simeq
1$. Note that, in principle, space-based data should perform better
than ground-based data, although a quantitative analysis of
space-based over ground-based performance at these redshifts remains
to be done under realistic observing conditions.

The alternative mass estimate we propose is based on lensing
magnification instead of shear. Magnification relies solely on
photometry instead of shapes, therefore any practical requirement on
how large a source galaxy must be vanishes; technically one could
measure lensing magnification on unresolved galaxies, well beyond
the redshift limits imposed on shape measurement by the PSF.
Magnification behind dark matter halos was pioneered by
\citet{broad1995,taylor1998}, but at that time, optical surveys were
too small and shallow, severely limiting photometric redshifts
estimates. Moreover, the net loss in signal-to-noise with
magnification compared to shear made the former less interesting for
practical applications \citep{ske2000}.

On the other hand, recent wide and deep surveys are well suited for
measuring the magnification signal. The measurement of cosmic
magnification on Lyman Break Galaxies \citep{hilde2009b} already
demonstrates that the weak lensing signal on $z\ge 2.5$ galaxies is
detectable. This paper explores this new opportunity for the study
of cluster/group-sized dark matter halos at high redshift.

Section 2 introduces the magnification technique, Section 3 uses the
population of high redshift galaxies detected in the
Canada-France-Hawaii Telescope Legacy Survey (CFHTLS) to demonstrate
how accurately halo masses and shapes can be measured. The last
Section discusses the prospects for this new technique in light of
current and forthcoming surveys.

\section{Galaxy-halo lensing magnification}

The gravitational lensing of the source galaxies is described by the
amplification matrix, ${\cal A}$, which is a function of shear,
$\gammag $, and convergence, $\kappa$ (see \cite{munshi2008} for a
review). The magnification, $\mu(\thetag)$, along a line-of-sight
quantifies the change in flux of a distant object affected by gravitational lensing:

\begin{equation}
\mu(\thetag)={1\over (1-\kappa(\thetag))^2-\gamma(\thetag)^2}
\label{mueq}
\end{equation}
where $\thetag=(\theta_1,\theta_2)$ is the position angle on the sky
measured from a chosen reference point. The apparent magnitude, $m$,
of a lensed object is $m+2.5 \log(\mu)$, making it brighter
($\mu>1$) or fainter ($\mu < 1$) depending on the amount of
projected mass present in the direction $\thetag$, as indicated in
equation \ref{mueq}. The magnification effect is measured by
counting, for lines-of-sight at different $\thetag$, the number of
galaxies that appear in a magnitude bin $[m,m+dm]$ for a survey of
limiting magnitude $m_{\rm lim}>m$. The number of unlensed galaxies
in $[m,m+dm]$ is $N_0(m)dm$ and the number of lensed galaxies in the
same magnitude range is $N(m,\thetag)dm$. \cite{n1989} showed that:

\begin{equation}
N(m,\thetag)dm=\mu^{2.5 s(m)-1}N_0(m)dm, \label{lensedcounts}
\end{equation}
where $s(m)$ is the slope of the logarithmic galaxy number counts at
magnitude $m$.
%
%
The magnification, $\mu$, can be directly obtained from a
measurement of the galaxy counts contrast, $\delta_N(\thetag)$:

\begin{equation}
\delta_N(\thetag)={N(m,\thetag)-N_0(m)\over N_0(m)}=\mu^{2.5
s(m)-1}-1 \label{numdef}
\end{equation}
\indent Behind a dark matter halo, galaxies are expected to be
magnified because both $\kappa$ and $\gamma$ are positive. A halo
could also be located in an under-dense region, leading to a dimming
of the background galaxies locally; however, with the average
lensing effect over the entire sky being zero, this environmental
effect vanishes and only contributes to the noise. Therefore, we can
safely focus on the mass profile of a halo itself, embedded in a
uniform background.

The strategy adopted in this work is to evaluate the constraints on
dark matter halos described by the universal Navarro-Frenk-White
(NFW) profile, $\rho_{\rm NFW}$ \citep{nfw1997}. The mass density
profile is described by the simple formula:

\begin{equation}
{\rho_{\rm NFW}(r) \over \rho_{\rm crit}}={\delta_0\over
(r/r_s)(1+r/r_s)^2}
\label{nfwdef}
\end{equation}
where $\rho_{\rm crit}$ is the critical density, $r_s$ is a
characteristic scale radius and $\delta_0$ is the density parameter:

\begin{equation}
\delta_0={200\over 3} {c_{200}^3\over \ln
(1+c_{200})-c_{200}/(1+c_{200})}
\end{equation}
The concentration parameter, $c_{200}$, at the virial radius,
$r_{200}$, is simply given by $c_{200}=r_{200}/r_s$. The parametric
model in equation \ref{nfwdef} effectively depends on only two
parameters: the concentration, $c_{200}$, and the halo mass,
$M_{200}$, both within the virial radius and related by
$M_{200}=4/3\pi r_{200}^3 \rho_{\rm crit}$. The velocity dispersion
is defined as $V_{200}=\sqrt{G M_{200}/r_{200}}$.

$\kappa$ and $\gamma$ are given by the second order derivatives of
the gravitational lensing potential, $\psi(\thetag)$:

\begin{eqnarray}
\kappa(\thetag)&=&{1\over 2}\left(\psi_{11}+\psi_{22}\right)
\nonumber \cr \gamma_1(\thetag)&=&{1\over
2}\left(\psi_{11}-\psi_{22}\right)\nonumber \cr
\gamma_2(\thetag)&=&\psi_{12}
\end{eqnarray}
$\psi(\thetag)$ essentially measures the projected mass along the
line-of-sight at $\thetag$:

\begin{equation}
\psi(\thetag)={4G\over c^2} {D_{ol} D_{os}\over D_{ls}}\int
d^2\theta'~\Sigma_{NFW}(\thetag') \ln |\thetag-\thetag'|
\end{equation}
where $\Sigma_{NFW}(\thetag)$ is the projected mass of the NFW
profile, and $D_{ol}$, $D_{os}$, $D_{ls}$ are the angular diameter
distances between the {\it source}, {\it observer} and the {\it
lens}.
%
%
%
Analytical expression of the lensing potential for the NFW profile
can be found in \cite{mene2003}. One could in principle constrain
the parameters $c_{200}$ and $V_{200}$ for each halo where the
magnification is measured. In practice however, the signal-to-noise
per halo is too low and it is necessary to stack the signal from
many foreground halos, a strategy similar to galaxy-galaxy lensing.
This technique is described in detail in the following section.

\section{Halo characterization}

Galaxy-galaxy lensing was initially developed as a probe of
galactic-size dark matter halos, making use of the tangential shear
around foreground galaxies \citep{brainerd1996,hudson1998}. Similar
to the shear, the signal-to-noise ratio of the lensing magnification
is low for most lenses, making the stacking of foreground lenses
necessary in order to lower the noise of the average magnification
as a function of distance from the lenses centers. The stacking
should rely on a {\it proxy} to group foreground lenses with similar
mass (e.g. stellar mass). It can be applied to clusters, groups of
galaxies and individual galaxies\footnote{Note that the most massive
clusters generate a strong magnification signal where stacking is
not necessary \citep{taylor1998}.}.
Constraints on the mass and concentration parameters, $M_{200}$ and
$c_{200}$, of the average magnification profile are obtained from
the likelihood:

\begin{equation}
{\cal L}\propto \exp
\left[(\delta_N\theta)-\bar\delta_N(\theta))C^{-1}_{{\delta_N}{\delta_N}}(\delta_N(\theta)-
\bar\delta_N(\theta))^{\rm\bf
T}\right]
\end{equation}

\noindent where $\delta_N(\theta)$ is the galaxy count profile from
equation \ref{numdef}, circularly averaged for our purpose, and
$\bar\delta_N(\theta)$ is the galaxy count profile model. The
covariance matrix, $C_{{\delta_N}{\delta_N}}$, is estimated as
described below.

In \cite{taylor1998}, it was shown that for a given population of
source galaxies, the net signal-to-noise loss for magnification is
larger by roughly a factor of $5$ compared to shear measurements.
Ignoring the sampling variance and the effect of large scale
structures, there are two sources of noise for magnification; 1) the
statistical, Poisson, noise due to discrete sampling of the source
galaxies and 2) the clustering of the source galaxies leading to
variations of number counts coherent over large angular distances.
In order to test the feasibility of our approach under realistic
observing conditions, the sampling and clustering sources of noise
in $C_{{\delta_N}{\delta_N}}$ are measured from a real population of
source galaxies. This population of source galaxies is the high
redshift, $z\ge 2.5$, Lyman Break Galaxies (LBGs) selected from the
CFHTLS-Wide data by the method described in \cite{hilde2009a}; it
contains $\sim130\,000$ $u$- and $g$- dropouts in the magnitude
range $23.5<i<24.5$ selected from 156 sq. deg. of the CFHTLS-Wide
dataset.

\begin{onecolumnfigure}
\begin{tabular}{c}
{\epsfxsize=7.cm\epsffile{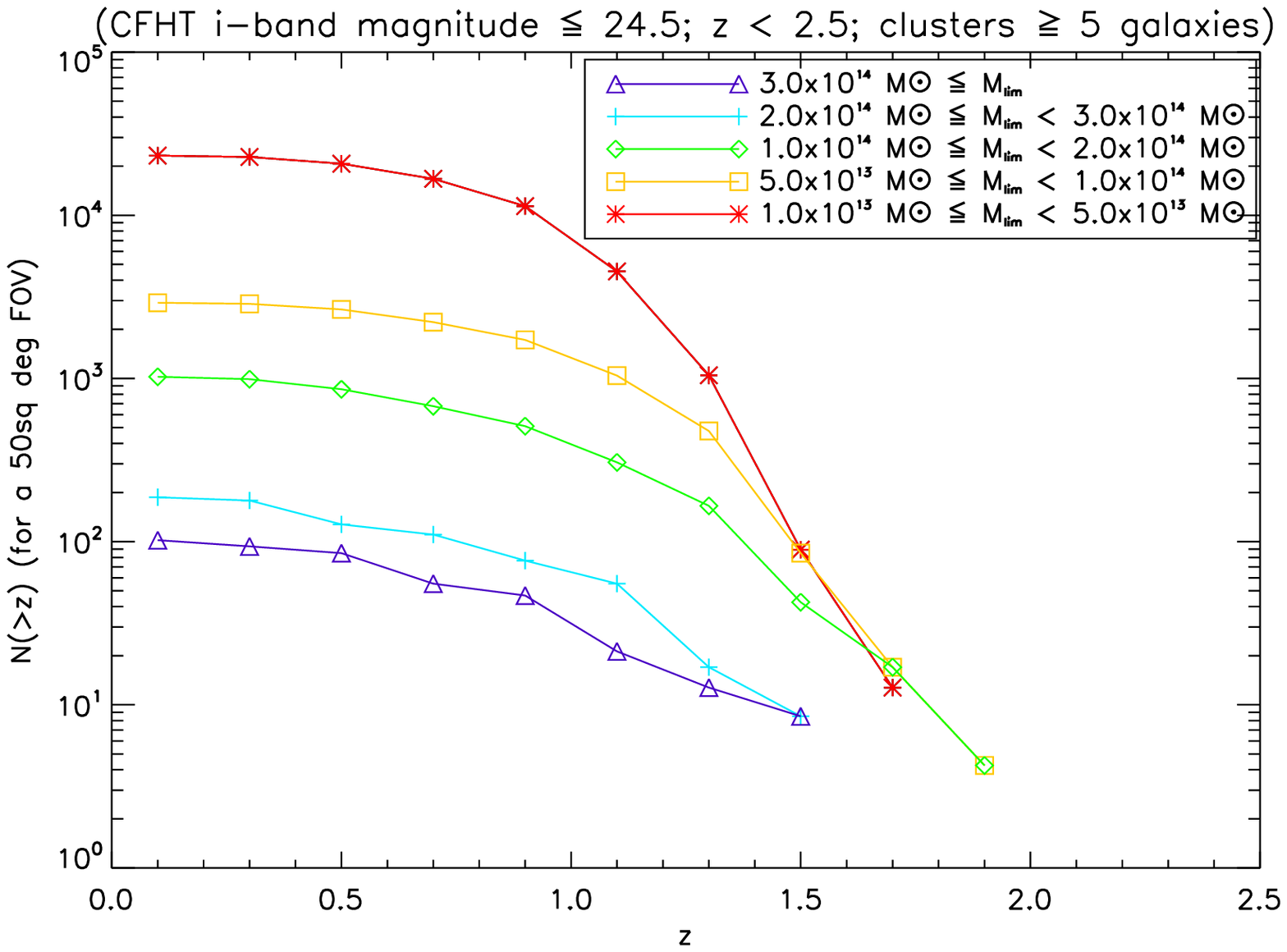}}
\end{tabular}
\figcaption{Cluster number counts from the Millennium Simulation
mock light cones of \cite{MSkitzbichler2007} as a function of
redshift for different mass bins. The counts have been normalized to
a $50$ sq. deg. survey. The depth of the galaxy catalogue matches
the CFHTLS-Wide data with $i<24.5$. Clusters are defined as detailed
in the text. \protect{\label{mscounts}}}
\end{onecolumnfigure}

\noindent   The slope of the number counts is such that
$2.5~s(m)-1=1.4$, as measured on the CFHTLS-Deep data with a
limiting magnitude approximately 2 magnitudes deeper than the
CFHTLS-Wide \citep{hilde2009b}.

The estimate of the noise covariance matrix,
$C_{{\delta_N}{\delta_N}}$, proceeds as follows: approximately $200$
random positions are chosen to represent foreground lenses in each
square degree of the 156 sq. deg. Their angular positions are then
cross-correlated to the LBGs for each square degree. The average
cross-correlation is zero, and the dispersion around zero
corresponds to the $C_{{\delta_N}{\delta_N}}$ expected for $200$
foreground lenses. The amplitude of the noise covariance matrix is
later adjusted to the actual number of stacked halos. Our procedure
takes into account realistic sampling and clustering noise of the
source population.

The realistic number of dark halos to stack for a given mass was
taken from the mock light cones of \cite{MSkitzbichler2007}, which
were created from the Millennium Simulation \citep{MSspringel2005}
derived galaxy catalogues of \cite{MSdelucia2007}.  Clusters in
these light cones are defined according to \cite{Milkeraitis2010}:
cluster members are identified as having the $M_{200}$ flag of their
parent halo, as well as the same friends-of-friends identification
number to avoid dark matter substructures.  Clusters are required to
have at least $5$ galaxy members and the light cones are filtered
out to a limiting magnitude of $i_{\mathrm{MegaCam}} = 24.5$, to be
in line with the CFHTLS data discussed herein.

Figure \ref{mscounts} shows the number of clusters per redshift for
a ground-based survey of 50 sq. deg.: for redshifts $>1$ we expect
$\sim 10^4$ clusters of mass $\sim 10^{13} M_{\odot}$ and $\sim 150$
clusters of mass $> 10^{14} M_{\odot}$. Note that the power spectrum
normalisation in the Millennium Simulation (MS) is $\sigma_8=0.9$
and thus substantially larger than the WMAP7 value, $\sigma_8=0.8$
\citep{larson2010}; therefore, the MS over-predicts the number of
halos by a factor of approximately $3-4$. The cluster number counts
shown in Figure \ref{mscounts} are thus approximately the expected
number in our Universe for a survey of $\sim 200$ sq. deg. with the
CFHTLS-Wide depth.

\begin{figure*}[t]
\hspace*{1cm}
\begin{tabular}{cc}
\psfig{file=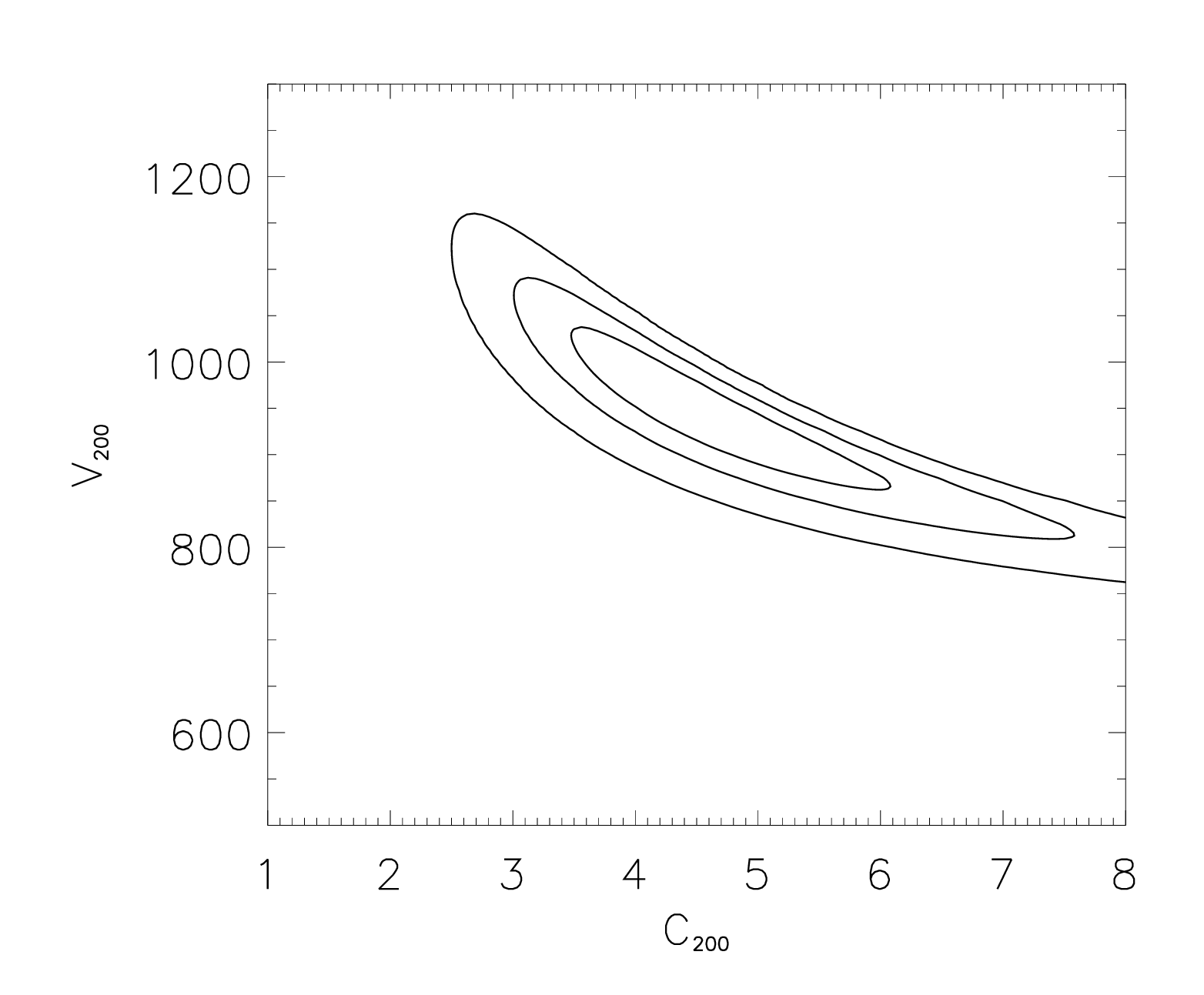,width=0.35\textwidth,clip=} &
\psfig{file=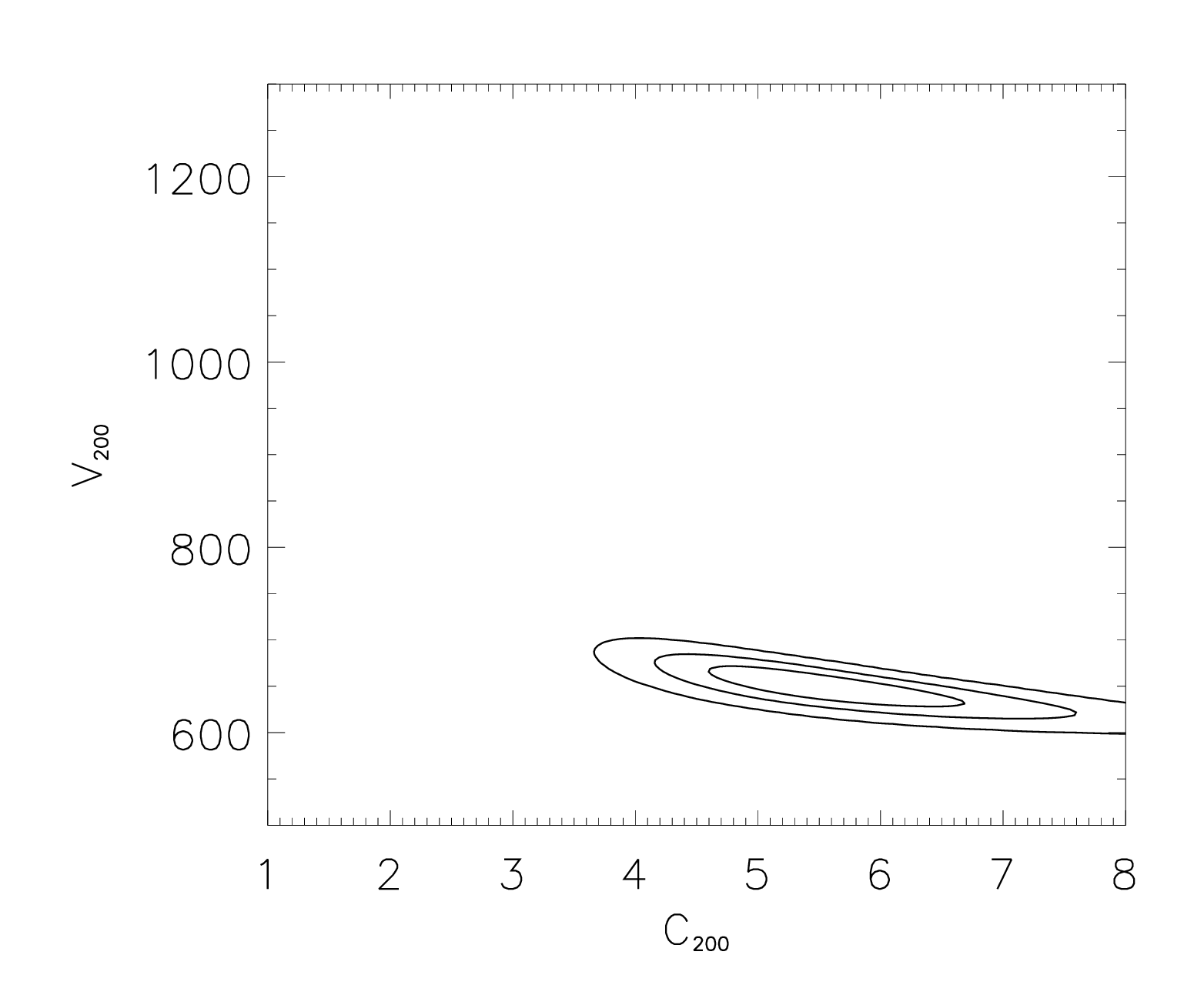,width=0.35\textwidth,clip=} \cr
\psfig{file=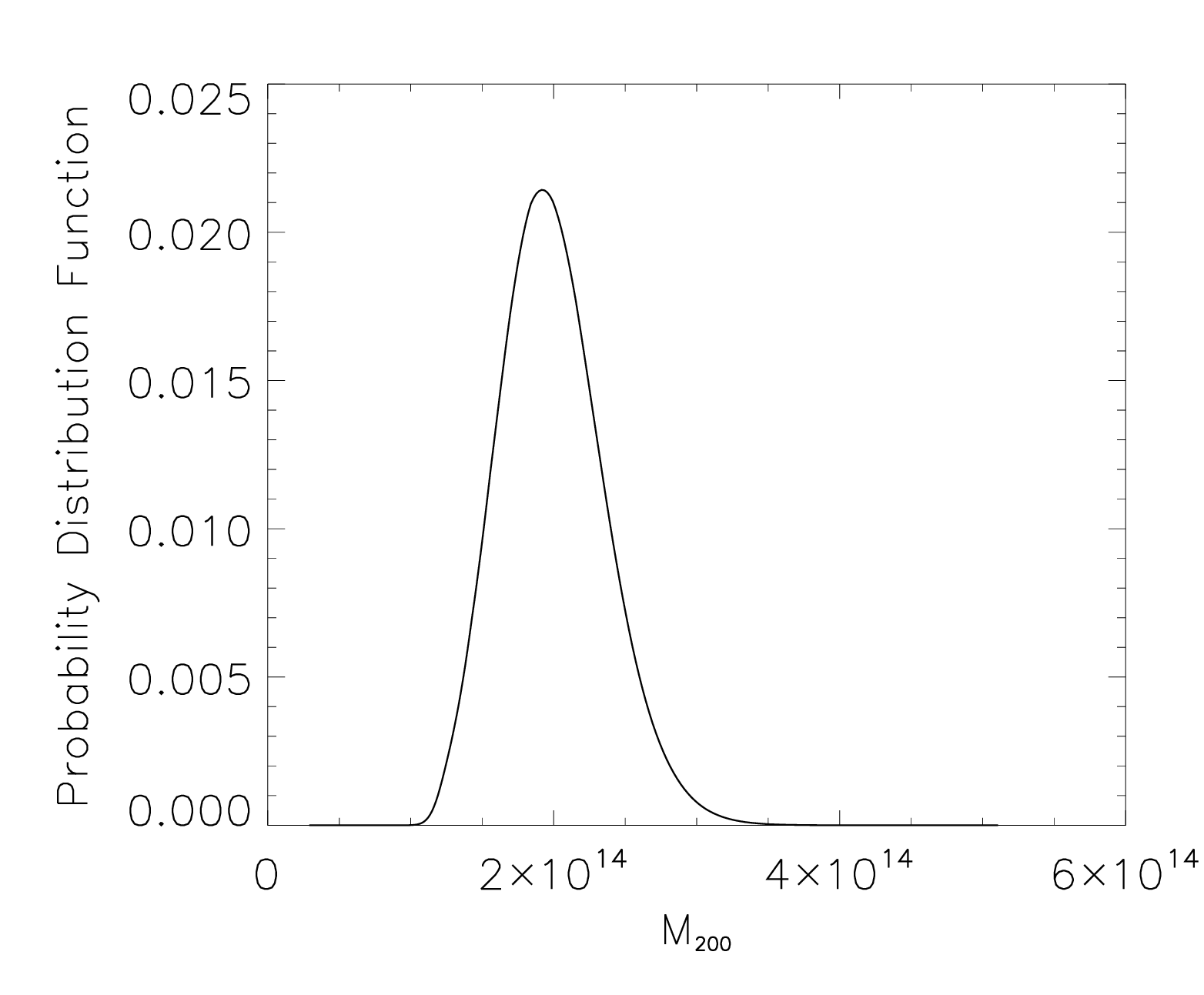,width=0.35\textwidth,clip=} &
\psfig{file=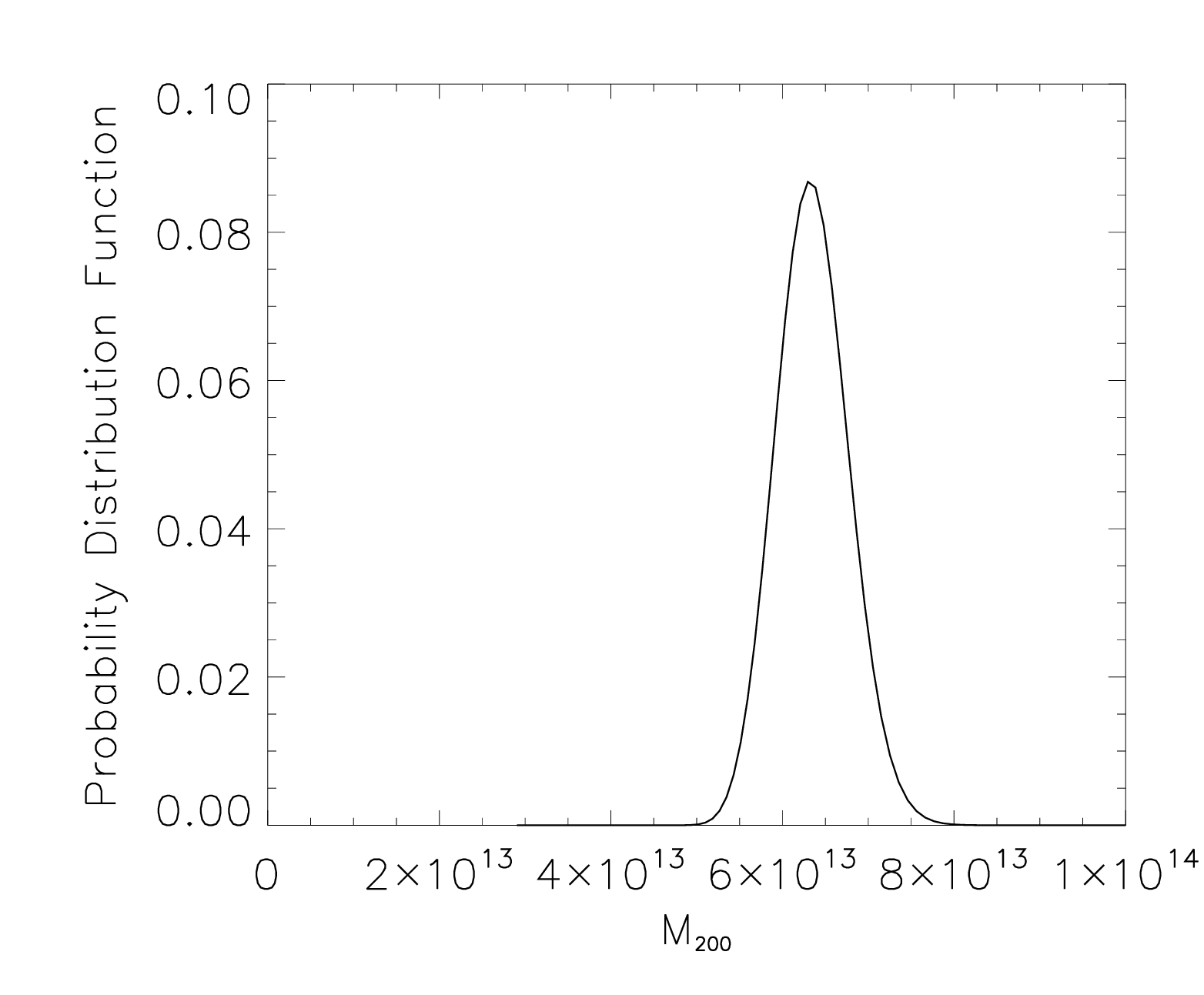,width=0.35\textwidth,clip=}
\end{tabular}
\caption{Top panels: the left plot shows the velocity dispersion,
$V_{200}$, versus concentration parameter, $c$: constraints from the
stacking of $135$ halos with $V_{200}=950~{\rm km/s}$ and
$c_{200}=4.5$ at redshift $z=1$; on the right, the constraints from
the stacking of $6000$ halos with $V_{200}=650~{\rm km/s}$ and
$c_{200}=5.5$ are shown. Bottom panels: left and right show the halo
mass derived from the constraints ($V_{200}$,$\ c$) from the top
left and top right panels respectively. \protect{\label{contours}}}
\end{figure*}

The performance of the magnification measurement is evaluated for a
low and a high mass bin, using halo counts taken from Figure
\ref{mscounts}. The low mass bin consists of $6000$ stacked halos
with $V_{200}=650~{\rm km/s}$ and $c_{200}=5.5$ and the high mass
sample contains $135$ stacked halos with $V_{200}=950~{\rm km/s}$
and $c_{200}=4.5$. All halos are placed at redshift $z=1$ and we
assume no centroid misalignment in the stacking. Figure
\ref{contours} shows the constraints on $V_{200}$ and $c_{200}$ for
mass bins using the magnification measured on a source of LBGs as
defined previously, assuming they are all at redshift $z=3$. For a
fixed concentration, the velocity dispersion for both mass bins is
constrained with an accuracy better than $5\%$. This shows that
magnification of high redshift galaxies can in principle probe dark
halos well beyond the domain of applicability of the shear method.
By extrapolating, it is clear that precision measurement of dark
halo mass profiles is possible with future full sky lensing surveys.

The practical implementation of this technique will require a mass
proxy and an estimate of the cluster/group center which coincides
with the halo peak. The mass proxy is not specified at this stage,
but it has to rely on indicators such as the stellar mass, X-ray
temperature/luminosity, cluster richness, or a combination of these.
Interestingly, the dispersion of the mass-richness relation
\citep{hilbert2010} appears significantly larger than the velocity
dispersion shown in Figure \ref{contours}. This strongly suggests
that the calibration of halo richness as a mass proxy using lensing
magnification can be refined for even smaller stacked foreground
samples. A centroid misalignment would cause a spread in the mass
profile, as suggested in \cite{mandelbaum2006} and
\cite{sheldon2004}. A complete study of the impact of center
misidentification is beyond the scope of this paper. Note that for
the two extremes of halo sizes, individual galaxies and massive
clusters, the centroiding issue is nearly nonexistent.

An interesting aspect of magnification is the possibility of
simultaneously measuring the lensing signal and the dust absorption
by the lenses. Dust absorption by the lens induces a small chromatic
angular scale-dependent cross-correlation between the background
population and the lens (unlike lensing magnification which is
achromatic).  \cite{menard2009a,menard2009b} showed that the dust
absorption is a very negligible source of magnitude noise and that
its contamination is only a few percent of the lensing signal. Only
small angular scales are affected, where baryonic matter is
concentrated and dust absorption is highest. The practical
implementation of how dust absorption can be fully integrated into a
magnification based mass measurement is left for a forthcoming
study.

\section{Conclusion}

We have shown that weak lensing magnification of $z\ge 2.5$ Lyman
Break Galaxies can reach a relatively high signal-to-noise to enable
the study of dark matter halos at high redshift, $z\ge 1$, where
traditional shear measurements would fail because the source
galaxies are unresolved. The performance of our method was
quantified using a real LBGs distribution, from the CFHTLS-Wide
data, as source galaxies. The steep number counts and the high
redshift of the sources considerably help alleviating the low
signal-to-noise found in previous magnification studies. The
approach is similar to galaxy-galaxy lensing, where lenses are
stacked according to a mass proxy, although no attempt was made to
discuss which mass proxy should be used since it entirely depends on
the data wavelength coverage.

The magnification technique as a probe of dark matter halos can be
generalized to a larger sample of background galaxies over a wider
magnitude and redshift range.
It is interesting to note that, in principle, the halo shape can
also be measured, which can be used to discriminate  between various
Cold Dark Matter models. The main limitations of our method lie in
our ability to separate lensing from dust extinction in large
surveys, which is dependent on the number and wavelength coverage of
filters, and in identifying reliable mass proxies and determining
the halo centers, although the latter two equally affect shear based
mass profile measurements, it is not specific to magnification. A
quantitative estimate of the practical limitations for future
lensing surveys such as LSST, and JDEM would be particularly
interesting, but is beyond the scope of this work.

\acknowledgements We are grateful to Peter Schneider for useful
comments on the manuscript. LVW and MM are supported by NSERC and
CIfAR. HH is supported by the European DUEL RTN project
MRTN-CT-2006-036133. JF is supported by a JPL grant number
$1394704$. The Millennium Simulation databases used in this paper
and the web application providing online access to them were
constructed as part of the activities of the German Astrophysical
Virtual Observatory.


\begin{thebibliography}{17}
\expandafter\ifx\csname
natexlab\endcsname\relax\def\natexlab#1{#1}\fi
\bibitem[{Allen} et~al.(2004){Allen}, {Schmidt}, {Ebeling}, {Fabian}, and {van Speybroeck}]{allen2004}
{Allen} S.W., et al., 2004, MNRAS, 353, 457

\bibitem[{Brainerd} et~al.(1996){Brainerd}, {Blandford}, and {Smail}]{brainerd1996}
{Brainerd} T.G., {Blandford} R.D., and {Smail} I., 1996, ApJ, 466, 623

\bibitem[{Broadhurst} et~al.(1995)]{broad1995}
{Broadhurst} T., {Taylor} A., and {Peacock} J., 1995, ApJ, 438, 49

\bibitem[{De Lucia} and {Blaizot}(2007)]{MSdelucia2007}
{De Lucia} G., and {Blaizot} J., 2007, MNRAS, 375, 2

\bibitem[{Hilbert} and {White}(2010)]{hilbert2010}
{Hilbert} S., and {White} S.D.M., 2010, MNRAS, in press, 289

\bibitem[{Hildebrandt} et~al.(2009{\natexlab{a}}){Hildebrandt}, {Pielorz}, {Erben}, {van Waerbeke},
 {Simon}, and {Capak}]{hilde2009a}
{Hildebrandt} H., {Pielorz} J., {Erben} T., et al.,
2009{\natexlab{a}}, A\&A, 498, 725

\bibitem[{Hildebrandt} et~al.(2009{\natexlab{b}}){Hildebrandt}, {van Waerbeke}, and {Erben}]{hilde2009b}
{Hildebrandt} H., {van Waerbeke} L., and {Erben} T., 2009{\natexlab{b}}, A\&A, 507, 683

\bibitem[{Hoekstra} et~al.(2005){Hoekstra}, {Hsieh}, {Yee}, {Lin}, and {Gladders}]{hoekstra2005}
{Hoekstra} H., {Hsieh} B.C., {Yee} H., et al., 2005, ApJ, 635, 73

\bibitem[{Hudson} et~al.(1998){Hudson}, {Gwyn}, {Dahle}, and {Kaiser}]{hudson1998}
{Hudson} M.J., {Gwyn} S., {Dahle} H., and {Kaiser} N., 1998, ApJ,
503, 531

\bibitem[{Kitzbichler} and {White}(2007)]{MSkitzbichler2007}
{Kitzbichler} M.G., and {White} S.D.M., 2007, MNRAS, 376, 2

\bibitem[{Larson} et~al.(2010){Larson}, {Dunkley}, {Hinshaw}, {Komatsu}, {Nolta},
{Bennett}, {Gold}, {Halpern}, {Hill}, {Jarosik}, {Kogut}, {Limon}, {Meyer}, {Odegard},
{Page}, {Smith}, {Spergel}, {Tucker}, {Weiland}, {Wollack}, and {Wright}]{larson2010}
{Larson} D., {Dunkley} J., {Hinshaw} G., {et al.}, 2010,
arXiv:1001.4635

\bibitem[{Mandelbaum} et al. (2006)]{mandelbaum2006}
{Mandelbaum} R., et al., 2006, MNRAS, 372, 758

\bibitem[{M{\'e}nard} et~al.(2009{\natexlab{a}}){M{\'e}nard}, {Kilbinger}, and {Scranton}]{menard2009a}
{M{\'e}nard} B., {Kilbinger} M., and {Scranton} R., 2009{\natexlab{a}}, MNRAS submitted, arXiv:0903.4199

\bibitem[{M{\'e}nard} et~al.(2009{\natexlab{b}}){M{\'e}nard}, {Scranton}, {Fukugita}, and {Richards}]
{menard2009b}
{M{\'e}nard} B., {Scranton} R., {Fukugita} M., and {Richards} G., 2009{\natexlab{b}}, MNRAS submitted,
 arXiv:0902.4240

\bibitem[{Meneghetti} et~al.(2003){Meneghetti}, {Bartelmann}, and {Moscardini}]{mene2003}
{Meneghetti} M., {Bartelmann} M., and {Moscardini} L., 2003, MNRAS, 340, 105

\bibitem[{Milkeraitis} et~al.(2010){Milkeraitis}, {Van Waerbeke}, {Heymans}, {Hildebrandt}, {Dietrich}, and
 {Erben}]{Milkeraitis2010}
{Milkeraitis} M., {Van Waerbeke} L., {Heymans} C., {Hildebrandt} H., {Dietrich} J.P., and {Erben} T., 2010,
 MNRAS accepted, arXiv:0912.0739

\bibitem[{Munshi} et~al.(2008){Munshi}, {Valageas}, {van Waerbeke}, and {Heavens}]{munshi2008}
{Munshi} D., et al., 2008, PhysRep, 462, 67

\bibitem[{Muzzin} et~al.(2009){Muzzin}, {Wilson}, {Yee}, {Hoekstra}, {Gilbank}, {Surace}, {Lacy},
{Blindert}, {Majumdar}, {Demarco}, {Gardner}, {Gladders}, and {Lonsdale}]{muzzin2009}
{Muzzin} A., {Wilson} G., {Yee} H.K.C., {et al.}, 2009, ApJ, 698, 1934

\bibitem[{Narayan}(1989)]{n1989}
{Narayan} R., 1989, ApJL, 339, 53

\bibitem[{Navarro} et~al.(1997){Navarro}, {Frenk}, and {White}]{nfw1997}
{Navarro} J.F., {Frenk} C.S., and {White} S.D.M., 1997, ApJ, 490, 493

\bibitem[{Schneider} et~al.(2000){Schneider}, {King}, and {Erben}]{ske2000}
{Schneider} P., {King} L., and {Erben} T., 2000, A\&A, 353, 41

\bibitem[{Sheldon} et~al.(2004)]{sheldon2004}
{Sheldon} E., et al., 2004, AJ, 127, 2544

\bibitem[{Springel} et~al.(2005){Springel}, {White}, {Jenkins}, {Frenk}, {Yoshida}, {Gao}, {Navarro},
{Thacker}, {Croton}, {Helly}, {Peacock}, {Cole}, {Thomas}, {Couchman}, {Evrard}, {Colberg}, and
{Pearce}]{MSspringel2005}
{Springel} V., {White} S.D.M., {Jenkins} A., {et al}, 2005, Nat, 435, 629

\bibitem[{Taylor} et~al.(1998){Taylor}, {Dye}, {Broadhurst}, {Benitez}, and {van Kampen}]{taylor1998}
{Taylor} A.N., et al., 1998, ApJ, 501, 539



\end{thebibliography}

\end{document}